\documentclass[twocolumn,english,aps,reprint]{revtex4-2}
\usepackage{lmodern}

\usepackage[T1]{fontenc}
\usepackage[latin9]{inputenc}
\usepackage{amsmath}
\usepackage{amssymb}
\usepackage{graphicx}
\usepackage{babel}
\begin{document}
\preprint{This line only printed with preprint option}
\title{Drive-Through Quantum Gate: Non-Stop Entangling a Mobile Ion Qubit
with a Stationary One}
\author{Ting Hsu$^{1,2,5}$, Wen-Han Png$^{3}$, Kuan-Ting Lin$^{5}$, Ming-Shien
Chang$^{4}$ and Guin-Dar Lin$^{1,2,5}$}
\affiliation{$^{1}$Department of Physics and Center for Quantum Science and Engineering,
National Taiwan University, Taipei 10617, Taiwan~\\
$^{2}$Physics Division, National Center for Theoretical Sciences,
Taipei 10617, Taiwan~\\
$^{3}$Centre for Quantum Technologies, National University of Singapore,
3 Science Drive 2, Singapore 117543~\\
$^{4}$Institute of Atomic and Molecular Sciences, Academia Sinica,
Taipei 10617, Taiwan~\\
$^{5}$Trapped-Ion Quantum Computing Laboratory, Hon Hai Research
Institute, Foxconn, Taipei 11492, Taiwan}
\begin{abstract}
Towards the scalable realization of a quantum computer, a quantum
charge-coupled device (QCCD) based on ion shuttling has been considered
a promising approach. However, the processes of detaching an ion from
an array, reintegrating it, and driving non-uniform motion introduce
severe heating, requiring significant time and laser power for re-cooling
and stabilization. To mitigate these challenges, we propose a novel
entangling scheme between a stationary ion qubit and a continuously
transported mobile ion, which remains in uniform motion and minimizes
motional heating. We theoretically demonstrate a gate error on the
order of $0.01\%$, within reach of current technology. This approach
enables resource-efficient quantum operations and facilitates long-distance
entanglement distribution, where stationary trapped-ion arrays serve
as memory units and mobile ions act as communication qubits passing
beside them. Our results pave the way for an alternative trapped-ion
architecture beyond the QCCD paradigm.
\end{abstract}
\maketitle
\textbf{Introduction.} Trapped ions are among the leading platforms
for quantum information processing \citep{Blatt2008,Monroe2013},
offering unparalleled coherence times \citep{Harty2014}, high-fidelity
gate operations \citep{Ballance2016,Gaebler2016,Srinivas2021}, and
straightforward qubit initialization and readout \citep{Harty2014}.
The inherent long-range Coulomb interaction enables all-to-all connectivity
in principle, allowing for scalable entanglement in linear ion chains
or two-dimensional ion lattices \citep{Wright2019,Wu2020,Mehdi2021}.
However, beyond hundreds of ions, the increasing number of vibrational
modes complicates mode addressing, laser control suffers from cross-talk
and challenges in maintaining individual ion addressing, and the interaction
strength between distant ions weakens significantly \citep{Zhang2017,Leung2018,Murali2020}.
These factors present fundamental challenges to scaling large trapped-ion
quantum processors. As system size continues to increase, preserving
these advantages while maintaining efficient and controllable connectivity
becomes a central challenge for trapped-ion architectures.

In response to these challenges, modular ion-trap architectures have
been proposed, leveraging either photonic interconnects \citep{Monroe2014,Stephenson2020,Krutyanskiy2023}
or ion shuttling \citep{Kielpinski2002,Wan2020}. While photonic links
enable entanglement over long distances, atom-photon interactions
are inherently weak and probabilistic, resulting in poor photon collection
efficiency and low success rates \citep{Bruzewicz2019}. The alternative
approach, shuttling-based architectures such as the quantum charge-coupled
device (QCCD), allows deterministic entangling operations by transporting
qubits into proximity through a sequence of transport steps \textendash{}
linear motion, merging, splitting, and swapping \citep{Pino2021,Moses2023}.
However, as quantum circuits grow in complexity, the number of transport
operations required in a QCCD increases rapidly, leading to cumulative
motional heating and decoherence \citep{Saki2022}. While laser and
sympathetic cooling can mitigate motional heating, these procedures
significantly extend operational time and increase laser power consumption
\citep{Pino2021}. These considerations indicate that transport-related
overhead can become a limiting factor in modular trapped-ion architectures,
particularly as the number of required transport operations increases.

\begin{figure*}
\includegraphics[width=2\columnwidth]{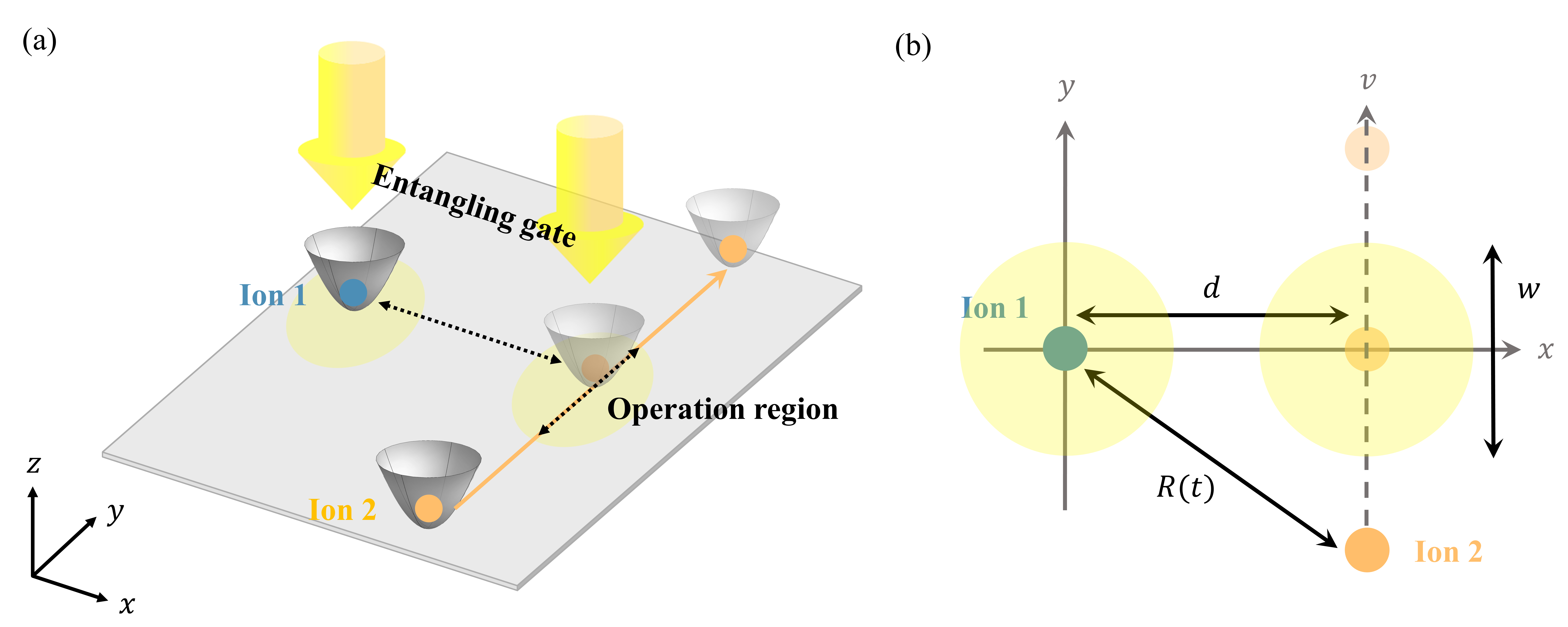}

\caption{\label{fig:1} \textbf{Drive-through gate scheme.} (a) Schematic representation
of the drive-through gate scheme for a two-ion system. The local trap
of ion $1$ is stationary while the local trap of ion $2$ is shuttled
along a straight path in the $x$-$y$ plane. The entangling operation
is performed as ion $2$ moves through the operation region. (b) Ion
$2$ undergoes uniform motion with speed $v$, leading to a time-dependent
Coulomb interaction characterized by $R(t)$, the inter-ion separation.
The laser beam, with a spatial width $w$, defines the operation region
where the drive-through gate is performed. }
\end{figure*}

To address these limitations, we propose a novel entangling scheme
with minimal transport overhead. As illustrated in Fig.~\ref{fig:1}a,
a moving ion, serving as a communication qubit, passes by a stationary
ion that acts as a memory qubit, while maintaining uniform motion
throughout the interaction. By incorporating the time-dependent Coulomb
interaction into the gate design, we theoretically demonstrate an
entangling gate with errors on the order of $10^{-4}$, achieved without
requiring additional cooling or stopping of the mobile qubit. The
same moving ion can subsequently interact with multiple stationary
qubits in a sequential manner, effectively functioning as a mobile
quantum bus for long-range entanglement generation. We refer to this
scheme as the drive-through gate (DTG).

Conventional trapped-ion architectures rely on entangling operations
performed at well-defined stopping points, where transported ions
are merged into stationary chains and subsequently re-separated. In
contrast, the DTG scheme removes the requirement that transport qubits
must be brought to rest for gate execution. Entanglement is instead
generated during continuous motion, with qubits remaining either stationary
or in uniform transport throughout the operation. While quantum gates
between co-moving ions have been demonstrated previously \citep{Tinkey2022,Leibfried2007},
such schemes rely on synchronized motion and do not address the scaling
overhead associated with repeated stopping, cooling, and reconfiguration
steps. By shifting the entangling interaction from discrete stopping
points to continuous transport, the DTG scheme defines a distinct
operational paradigm for modular distributed trapped-ion quantum computing,
enabling interactions between spatially separated memory units with
reduced transport-induced overhead.

\begin{flushleft}
\textbf{Results }
\par\end{flushleft}

\textbf{Setup.} To illustrate the DTG scheme, we consider a two-ion
system as shown in Fig.~\ref{fig:1}a. Ion 1 is confined by a stationary
harmonic trap whose center is fixed at $Q_{1}=(0,0,0)$ while ion
2 is captured by a moving harmonic trap whose center follows a prescribed
trajectory $Q_{2}(t)=(d,vt,0)$. In the presence of the Coulomb interaction,
the equilibrium position of each ion is shifted relative to the center
of its trapping potential. The actual position of the $i$-th ion
can therefore be written as $\boldsymbol{q}_{i}(t)=\boldsymbol{q}_{i}^{(0)}(t)+\boldsymbol{\xi}_{i}(t)$,
where $\boldsymbol{q}_{i}^{(0)}(t)$ denotes the instantaneous equilibrium
position and $\boldsymbol{\xi}_{i}(t)$ describes small motional oscillation
around it. In the adiabatic shuttling regime considered here, these
oscillations remain small compared to the inter-ion separation, allowing
the transverse motion to be treated independently from the in-plane
transport.

In the following, we focus on the transverse ($z$) motional modes
for gate implementation. This choice is motivated by experimental
considerations: residual acceleration arising from electrode noise
primarily affects in-plane transport, and laser coupling to in-plane
modes is more susceptible to Doppler shifts during shuttling. By operating
along the transverse direction, these effects are naturally suppressed,
allowing the gate dynamics to be analyzed within a more controlled
regime. This separation enables a time-dependent normal-mode description
that remains well-defined throughout the transport process.

\textbf{Dynamical motional normal modes.} In the adiabatic shuttling
regime, the motional oscillations $\boldsymbol{\xi}_{i}(t)$ around
the equilibrium positions satisfy $\left|\boldsymbol{\xi}_{i}(t)\right|\ll R(t)$,
where $R(t)=\left|\boldsymbol{q}_{1}^{(0)}(t)-\boldsymbol{q}_{2}^{(0)}(t)\right|$
denotes the time-dependent separation between the instantaneous equilibrium
positions of the two ions. Under this condition, the Coulomb interaction
can be expanded to second order in the transverse motion, leading
to an effective quadratic Hamiltonian $\hat{H_{0}}(t)=\sum_{i}\frac{\hat{p}_{i}^{2}}{2m}+\frac{1}{2}\sum_{i,j=1}^{2}G_{ij}\hat{z}_{i}\hat{z}_{j}$,
where $\hat{z}_{i}$ denotes the transverse position operator of the
$i$th ion measured from its instantaneous equilibrium position $\boldsymbol{q}_{i}^{(0)}(t)$,
and $G_{ij}=m\omega_{z}^{2}\delta_{ij}-\frac{K}{R^{3}(t)}(-1)^{i+j}$,
with $K=\frac{e^{2}}{4\pi\epsilon_{0}}$. 

Following the dynamical normal-mode formalism of Ref.~\citep{Lizuain2017},
because the two ions are of the same species and experience identical
local confinement, the transverse motion can be described in terms
of the dynamical normal modes despite the time-dependent Coulomb coupling.
The system admits a center-of-mass mode and a zigzag mode with time-dependent
frequencies 
\begin{eqnarray}
\Omega_{1}(t) & = & \omega_{z},\\
\Omega_{2}(t) & = & \sqrt{\omega_{z}^{2}-\frac{2K}{mR^{3}(t)}}.
\end{eqnarray}
In this basis, the Hamiltonian becomes diagonal, 
\begin{equation}
\hat{H}_{0}(t)=\frac{1}{2}\sum_{n=1}^{2}\left[\frac{\hat{P}_{n}^{2}}{m}+m\Omega_{n}^{2}(t)\hat{Z}_{n}^{2}\right],
\end{equation}
where $\hat{Z}_{n}$ and $\hat{P}_{n}$ are the canonical position
and momentum operators associated with the $n$th dynamical normal
mode.

\begin{figure*}
\includegraphics[width=2\columnwidth]{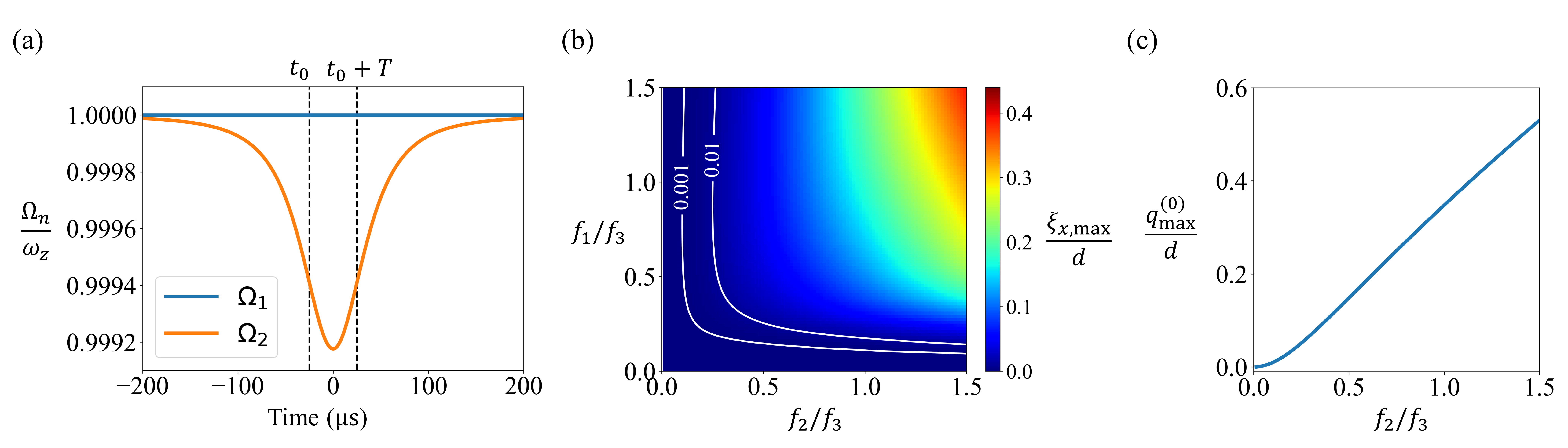}
\caption{\label{fig:2} \textbf{Dynamical normal modes and parameter regimes
for the DTG scheme.} (a) The dynamical normal mode frequencies. The
blue and orange curves correspond to the center-of-mass mode $\Omega_{1}$
and zigzag mode $\Omega_{2}$, respectively. The trapping frequency
is $\omega_{x,y}=2\pi\times2.5~{\rm MHz}$ and $\omega_{z}=2\pi\times5~{\rm MHz}$,
the closest trap distance is $d=10~\mu{\rm m}$, the shuttling velocity
is $v=0.2~{\rm m/s}$ and the laser addressing width is $w=10~\mu{\rm m}$.
The two black dashed lines indicate the entry and exit times of the
mobile ion within the operation region, corresponding to $t_{0}=-w/(2v)$
and $T=w/v$. (b) Heatmap showing the normalized oscillation amplitude
in the $x$-direction, $\xi_{x,{\rm max}}/d$, as a function of the
frequency ratios $f_{1}/f_{3}$ and $f_{2}/f_{3}$. The contours indicate
specific values of $\xi_{x,{\rm max}}/d$. (c) Maximum displacement
of the equilibrium positions of the ions, normalized as $q_{{\rm max}}^{(0)}/d$,
plotted as a function of $f_{2}/f_{3}$.}
\end{figure*}

\textbf{Gate design.} The drive-through gate is implemented using
a state-dependent force that couples the internal qubit states to
the transverse motional degrees of freedom. When the moving ion enters
the operation region, laser beams are simultaneously applied to both
ions to generate a spin-dependent displacement along the transverse
($z$) direction. 

In the Lamb\textendash Dicke regime, where the motional scales are
small compared to the optical wavelength, the laser-ion interaction
can be expressed, in the dynamical normal-mode basis, as $\hat{H}_{1}(t)=f(t)\sum_{j,n=1}^{2}b_{j}^{n}\hat{Z}_{n}\hat{\sigma}_{j}^{z}$.
Here, $\hat{\sigma}_{j}^{z}$ denotes the Pauli operator of the $j$th
ion, $\hat{Z}_{n}$ is the position operator of the $n$th dynamical
normal mode, and $b_{j}^{n}$ is the corresponding mode participation
coefficient. The time-dependent force amplitude is given by $f(t)=-\hbar k\chi(t)\sin\mu t$
\citep{GarciaRipoll2005,Zhu2006}, where $\chi(t)$ is the effective
Rabi frequency envelope, $k$ is the effective wavenumber of the laser
field, and $\mu$ is the detuning relative to the qubit transition.
The laser is applied only when the moving ion traverses a finite operation
region of width $w$ along its trajectory, such that $\chi(t)=0$
outside this interval. The total Hamiltonian of the system is therefore
$\hat{H}(t)=\hat{H}_{0}(t)+\hat{H}_{1}(t)$, which forms the starting
point for the derivation of the drive-through entangling gate.

DTG differs from conventional geometric phase gates in that the motional
mode frequencies $\Omega_{n}(t)$ vary in time during the interaction.
Nevertheless, the underlying mechanism is the same: a state-dependent
force induces conditional displacements in phase space, and a closed-loop
trajectory yields a qubit-state-dependent geometric phase.

To make this explicit, we perform a unitary transformation with the
unitary operator

\begin{eqnarray}
\hat{U}_{D} & = & \exp\left\{ \frac{i}{\hbar}\sum_{j,n=1}^{2}\left[m\dot{u}_{n}(t)\hat{Z}_{n}-u_{n}(t)\hat{P}_{n}\right]b_{j}^{n}\hat{\sigma}_{j}^{z}\right\} ,
\end{eqnarray}
 where $u_{n}(t)$ satisfies the classical equation of motion of a
driven oscillator with time-dependent frequency,
\begin{equation}
\ddot{u}_{n}(t)+\Omega_{n}^{2}(t)u_{n}(t)=\frac{f(t)}{m},\label{eq:classical EOM}
\end{equation}
together with initial conditions $u_{n}(t_{0})=0$ and $\dot{u}_{n}(t_{0})=0$
so that $\hat{U}_{D}(t_{0})=\hat{I}$. Under this transformation,
the Hamiltonian $\hat{H}(t)$ is mapped into $\tilde{H}(t)=\hat{U}_{D}\hat{H}(t)\hat{U}_{D}^{\dagger}-i\hbar\hat{U}_{D}\frac{\partial\hat{U}_{D}^{\dagger}}{\partial t}=\hat{H}_{0}(t)+\hat{\theta}(t),$
where $\hat{\theta}$ contains only spin operators (see Method), 
\begin{equation}
\hat{\theta}(t)=-\frac{1}{2}f(t)\sum_{j,l,n=1}^{2}u_{n}(t)b_{j}^{n}b_{l}^{n}\hat{\sigma}_{j}^{z}\hat{\sigma}_{l}^{z}.
\end{equation}
 As a result, the full evolution operator associated with $\hat{H}(t)$
can be factorized as 
\begin{eqnarray}
\hat{U}(t) & = & \hat{U}_{D}^{\dagger}(t)\mathcal{T}\exp\left[-\frac{i}{\hbar}\int_{t_{0}}^{t}\hat{\theta}(\tau)d\tau\right]\hat{U}_{0}(t),
\end{eqnarray}
where $\hat{U}_{0}(t)=\mathcal{T}\exp\left[-\frac{i}{\hbar}\int_{t_{0}}^{t}\hat{H}_{0}(\tau)d\tau\right]$
is the free evolution operator in the absence of laser driving.

We define $\hat{M}(t)$ as the operation on top of the free evolution,
which can be represented as \begin{widetext} 
\begin{equation}
\hat{M}(t)=\exp\left\{ -\frac{i}{\hbar}\sum_{j,n=1}^{2}\left[m\dot{u}_{n}(t)\hat{Z}_{n}-u_{n}(t)\hat{P}_{n}\right]b_{j}^{n}\hat{\sigma}_{j}^{z}\right\} \exp\left[\frac{i}{2\hbar}\int_{t_{0}}^{t}d\tau f(\tau)\sum_{j,l,n=1,j\neq l}^{2}u_{n}(\tau)b_{j}^{n}b_{l}^{n}\hat{\sigma}_{j}^{z}\hat{\sigma}_{l}^{z}\right]
\end{equation}
 \end{widetext} where we have ignored irrelevant phases. To ensure
at $t=t_{0}+T$ that $\hat{M}$ is a gate operator of the controlled-phase-flip
(CPF) gate, we require that 
\begin{eqnarray}
u_{n}(t_{0}+T) & = & 0,\label{eq: gate requirement for displacement}\\
\dot{u}_{n}(t_{0}+T) & = & 0,
\end{eqnarray}
 and 
\begin{equation}
\phi(t_{0}+T)=\frac{1}{2\hbar}\int_{t_{0}}^{t_{0}+T}d\tau f(\tau)\sum_{j,l,n=1,j\neq l}^{2}u_{n}(\tau)b_{j}^{n}b_{l}^{n}=-\frac{\pi}{4}.\label{eq: gate requirement for phase}
\end{equation}

\textbf{Constraints.} The derivation above relies on operating in
a regime where the dynamical normal-mode description remains valid
during transport. We now examine the corresponding constraints on
the system parameters. In our model, we assume that each ion is initially
at its equilibrium position within its local trap, so there is no
intrinsic oscillation due to the trapping potential. As the moving
ion approaches and interacts with the stationary ion via Coulomb forces,
the equilibrium positions of the two ions are displaced equally but
in opposite directions from their respective trap centers. Additionally,
the Coulomb interaction induces anti-symmetric oscillations of the
ions around their new, shifted equilibrium positions, with both ions
oscillating with equal amplitude but in opposite directions. 

The DTG scheme requires an adiabatic shuttling such that $\left|\boldsymbol{\xi}_{i}(t)\right|\ll R(t)$,
which imposes constraints on the choice of parameters $v$, $d$ and
$\omega_{x,y}$. This condition can be reformulated as $\xi_{\mu,{\rm max}}/d\ll1$,
where $\xi_{\mu,{\rm max}}$ represents the maximum oscillation amplitude
during the gate operation in the $\mu$-direction. We identify three
frequency scales that influence $\xi_{\mu,{\rm max}}/d$. The first
is $f_{1}=v/d$, which characterizes the dynamics of the shuttling
process. The second is $f_{2}=\sqrt{\frac{e^{2}}{4\pi\epsilon_{0}md^{3}}}$,
associated with the strength of the Coulomb interaction between the
two ions. The third is $f_{3}=\omega_{x,y}$, reflecting the strength
of the trapping potential. Fig.~\ref{fig:2}b illustrates the normalized
oscillation amplitude $\xi_{\mu,{\rm max}}/d$ as a function of the
frequency ratios $f_{1}/f_{3}$ and $f_{2}/f_{3}$. For simplicity,
only $\xi_{x,{\rm max}}$ is plotted, as $\xi_{x,{\rm max}}$ and
$\xi_{y,{\rm max}}$ are of comparable magnitude. In the strong trapping
regime, where $f_{1}/f_{3}\ll1$ or $f_{2}/f_{3}\ll1$, the confinement
provided by the trap effectively suppresses the ion's oscillatory
motion, ensuring that $\xi_{\mu,{\rm max}}/d\ll1$, satisfying the
adiabatic shuttling requirement. Conversely, when $f_{1}\sim f_{2}\sim f_{3}$,
the oscillations become more pronounced. 

We further examine the displacement of the equilibrium positions of
the ions caused by the Coulomb interaction. Fig.~\ref{fig:2}c shows
the normalized displacement $q_{{\rm max}}^{(0)}/d$ as a function
of $f_{2}/f_{3}$, where $q_{{\rm max}}^{(0)}$ represents the maximum
displacement during the gate operation. While it is theoretically
possible for a DTG to operate effectively even with large displacements,
we opt to perform operations in the small-displacement regime, where
$q_{{\rm max}}^{(0)}/d\ll1$. Operating in this regime offers several
advantages. First, during gate operation, laser beams are used to
drive the interaction, and the laser intensity is typically strongest
and most robust at the beam center. By minimizing the displacement,
the ions remain closer to the center of the laser field, thereby reducing
the required laser power and undesired fluctuation. Second, a smaller
displacement ensures greater stability within the trap. Displacement
from the trap minimum introduces greater micromotion for the ions,
which can degrade the fidelity of the gate operation. By confining
the ions to a small-displacement regime, we enhance the robustness
of the gate while optimizing the power efficiency of the system.

\textbf{Pulse shaping.} In an ideal geometric-phase gate, one would
satisfy the closure conditions $u_{n}(t_{0}+T)=0$ and $\dot{u}_{n}(t_{0}+T)=0$
for all motional modes while accumulating the target entangling phase
$\phi(t_{0}+T)=-\pi/4$. In practice, the time-dependent mode frequencies
$\Omega_{n}(t)$ and the finite parametrization of the drive prevent
these constraints from being met exactly. We therefore quantify the
residual spin\textendash motion entanglement by the final phase-space
displacement, $\Delta u_{n}=u_{n}(t_{0}+T)$, $\Delta\dot{u}_{n}=\dot{u}_{n}(t_{0}+T)$
and the phase mismatch $\Delta\phi=\phi(t_{0}+T)+\pi/4$. To leading
order in these residuals, the gate fidelity can be expressed as 
\begin{equation}
F=1-\frac{m}{2\hbar}\sum_{n=1}^{2}\left(\frac{|\Delta\dot{u}_{n}|^{2}}{\omega_{z}}+\omega_{z}|\Delta u_{n}|^{2}\right)\left(2\bar{N_{n}}+1\right)-|\Delta\phi|^{2},
\end{equation}
where $\bar{N_{n}}$ is the mean phonon occupation of mode $n$ in
the initial thermal state.

We optimize the gate performance by tailoring the amplitude envelope
$\chi(t)$ and the detuning $\mu$ of the driving field \citep{Choi2014}.
The drive is applied only while the moving ion traverses the operation
region, so that $\chi(t)=0$ outside the gate window, {[}$t_{0},t_{0}+T${]},
with $t_{0}=-w/(2v)$ and $T=w/v$. Within this interval, we parametrize
$\chi(t)$ as a piecewise-constant function consisting of five equal-duration
segments. The set of segment amplitudes $\left\{ \chi_{k}(t)\right\} $
and the detuning $\mu$ are then determined by numerical optimization
to maximize $F$ for a given experimental configuration.

Fig.~\ref{fig:3} shows two representative optimized solutions for
a pair of $^{171}\text{Yb}^{+}$ ions with $d=w=10$ $\mu$m. For
a shuttling velocity $v=0.2$ m/s, the optimal parameters are $\mu=-0.06\omega_{z}$,
and the corresponding five-segment envelope is shown in Fig.~\ref{fig:3}a;
the resulting phase-space trajectories of both dynamical modes return
to the origin with high accuracy (Fig.~\ref{fig:3}b). For a faster
transport, $v=0.5$ m/s, the optimal detuning shifts to $\mu=-0.02\omega_{z}$
with a modified envelope (Fig.~\ref{fig:3}c), again yielding near-closed
trajectories (Fig.~\ref{fig:3}d). In both cases, the optimized five-segment
drive achieves an intrinsic fidelity approaching $F\approx1-10^{-7}$
assuming initial thermal states at the Doppler limit.
\begin{flushleft}
\begin{figure*}
\includegraphics[width=2\columnwidth]{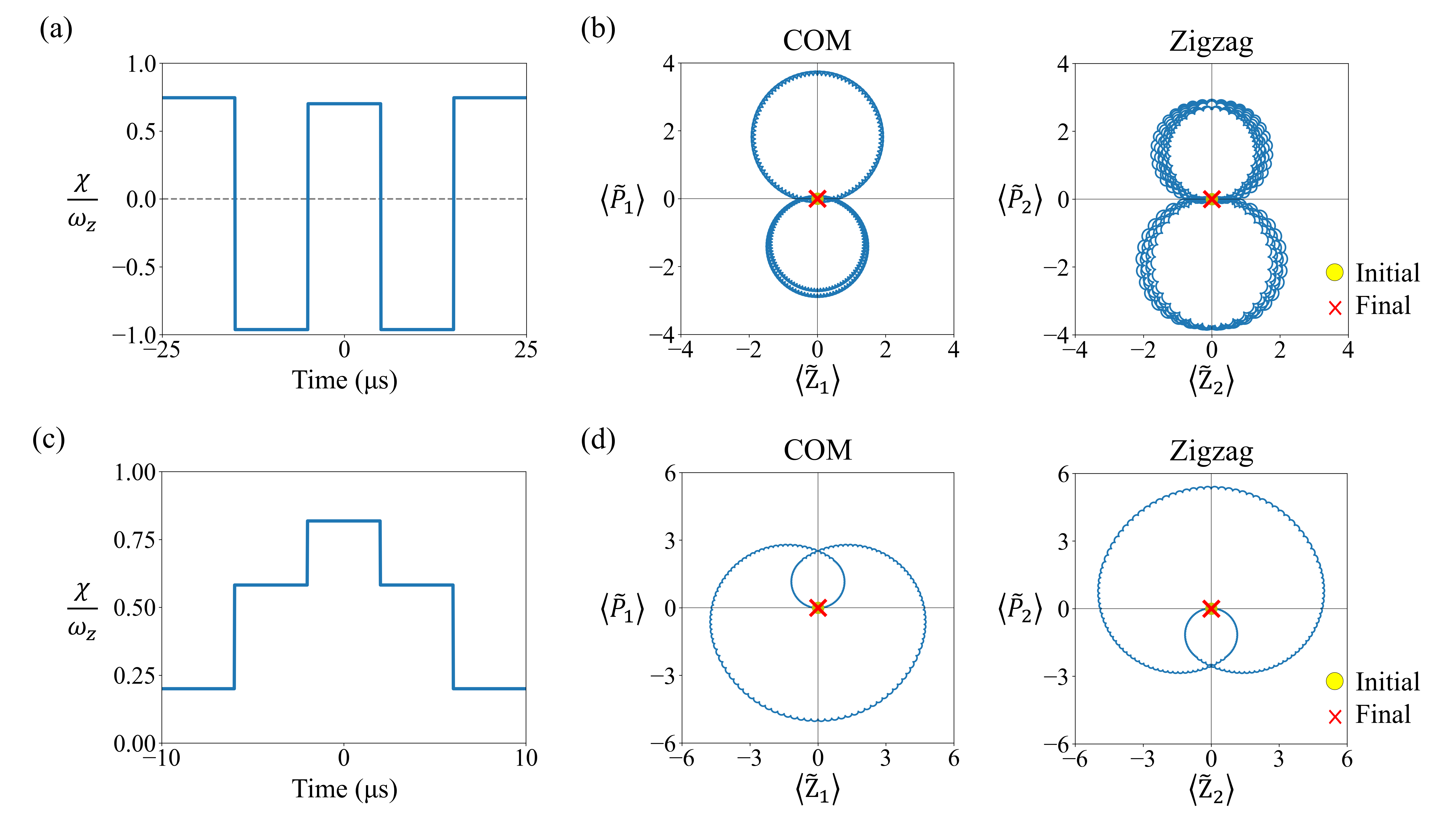} 

\caption{\label{fig:3}\textbf{ Optimized pulse shape and phase space trajectory.}
(a) Optimized pulse shape and (b) phase space trajectories with $v=0.2~{\rm m/s}$
and $\mu=-0.06\omega_{z}$. (c) Optimized pulse shape and (d) phase
space trajectories with $v=0.5~{\rm m/s}$ and $\mu=-0.02\omega_{z}$.
In both cases, $d=10~\mu{\rm m}$, $w=10~\mu{\rm m}$, $\omega_{x,y}=2\pi\times2.5~{\rm MHz}$
and $\omega_{z}=2\pi\times5~{\rm MHz}$. $\left\langle \tilde{Z}_{n}\right\rangle $
and $\left\langle \tilde{P}_{n}\right\rangle $ denote the normalized
expectation values of the position and momentum operators in the interaction
picture for the center-of-mass ($n=1$) and zigzag ($n=2$) modes.
In (b) and (d), the yellow circle (initial) and red cross (final)
indicate the phase-space displacements.}
\end{figure*}
\par\end{flushleft}

\textbf{Error analysis.} The intrinsic error of the drive-through
gate arises from residual spin\textendash motion entanglement due
to imperfect closure of the phase-space trajectories. With optimized
pulse shaping, this contribution can be suppressed to the level of
$\delta F_{1}\sim10^{-7}$, indicating that the protocol-level error
is negligible under idealized conditions.

A second class of errors is associated with the shuttling-induced
motion of the ions in the $x$-$y$ plane. Such motion leads to fluctuations
in the effective Rabi frequency experienced by the ions during the
interaction. For small oscillations in the $x$-$y$ plane, the resulting
infidelity scales as $\delta F_{2}\approx\left(\pi/2\right)\left(\xi_{\mu,{\rm max}}/w\right)^{4}$
\citep{Lin2009}, which remains below $10^{-10}$ within the parameter
regime considered. This contribution is therefore strongly suppressed
by operating in the small-displacement regime identified above.

Additional errors originate from standard experimental imperfections
common to trapped-ion platforms. In particular, higher-order terms
beyond the Lamb\textendash Dicke approximation for thermal motion
along the $z$ axis contribute an estimated infidelity of order $\delta F_{3}\approx10^{-4}$
for ions pre-cooled to the Doppler temperature \citep{Soerensen2000}.
Other effects, including crosstalk and laser intensity drift, can
be further suppressed using established pulse-shaping and compensation
techniques \citep{Bluemel2021}. The unwanted Stark shift associated
with the $\sigma_{z}\otimes\sigma_{z}$ interaction can be eliminated
by echo pulses \citep{Clark2021,Sawyer2021}. Although optical phase
noise is not treated explicitly, its effect can be mitigated using
established optimized control protocols \citep{Steane2014}.

\begin{figure*}
\includegraphics[width=2\columnwidth]{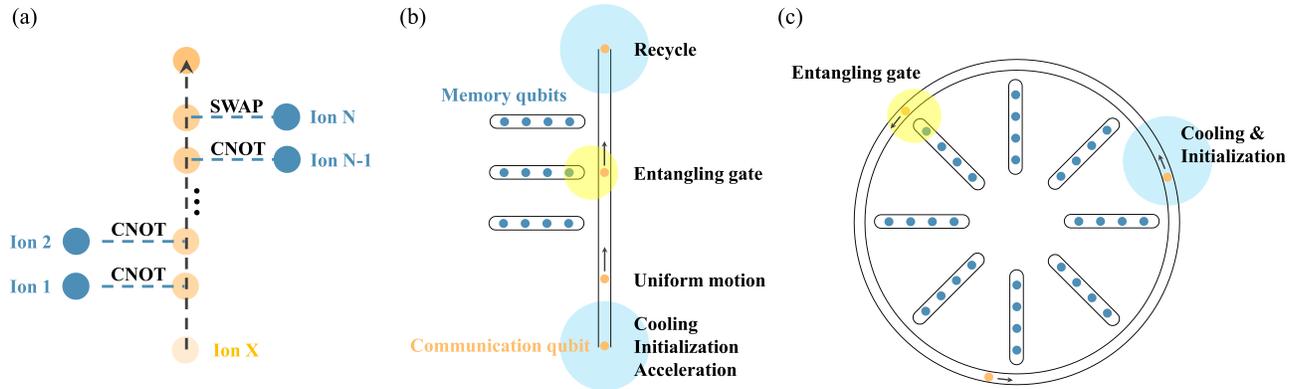}
\caption{\label{fig:4} \textbf{Scalable architecture.} (a) Schematic for generating
a distributed $N$-qubit GHZ state using a single moving ion and $N$
stationary ions. The moving ion (Ion $X$) is initialized to a superposition
state and sequentially interacts with stationary ions $1$ through
$N-1$ via drive-through CNOT gates. A final drive-through SWAP gate
with stationary ion $N$ completes the process, creating a GHZ state
distributed among the $N$ stationary ions. (b) Linear architecture
with static ion arrays serving as memory units. A single moving ion
is initialized in the cooling and initialization zone, interacts with
the memory ions via drive-through gates, and can be recycled afterward.
(c) Race-track architecture where multiple moving ions travel along
a closed track, interacting with memory ions via drive-through gates.
The moving ions can be re-cooled and re-initialized after completing
a cycle. }
\end{figure*}

\textbf{Scalable architecture.} Having established the feasibility
and error characteristics of the drive-through gate at the two-ion
level, we now discuss how this primitive can be incorporated into
scalable trapped-ion architectures. The key feature of the DTG scheme
is that entanglement is generated during continuous transport, allowing
moving ions to function as communication qubits that interact sequentially
with spatially separated stationary ions acting as memory units. As
an illustrative example, Fig.~\ref{fig:4}a shows how a distributed
$N$-qubit GHZ state $\left(|0\rangle^{\otimes N}+|1\rangle^{\otimes N}\right)/\sqrt{2}$
can be generated using a single moving ion and $N$ stationary ions.
The moving ion (Ion $X$) is first prepared in a superposition state
and subsequently undergoes a sequence of drive-through CNOT interactions
with stationary ions $1$ through $N-1$. A final drive-through SWAP
operation with stationary ion $N$ transfers the entanglement to the
stationary register, resulting in a GHZ state distributed across the
memory ions. This example highlights how a single mobile qubit can
mediate long-range entanglement without requiring repeated stopping,
merging, or reconfiguration steps.

Beyond this illustrative protocol, the drive-through gate naturally
supports modular architectures in which stationary ion arrays serve
as local memory units while moving ions provide inter-module connectivity.
Figures~\ref{fig:4}b and \ref{fig:4}c depict two representative
realizations of this concept. In the linear architecture shown in
Fig.~\ref{fig:4}b, a moving ion is initialized in a designated cooling
and preparation region, interacts sequentially with multiple memory
arrays via drive-through gates, and can be recycled after completing
its interactions. Fig.~\ref{fig:4}c illustrates a race-track architecture
in which multiple moving ions circulate along closed paths, periodically
passing through cooling and initialization zones. Such architectures
enable continuous operation and flexible routing of quantum information
while avoiding the transport overhead associated with repeated stopping
and re-cooling.

Importantly, the present work focuses on establishing the gate-level
physical mechanism underlying the drive-through paradigm. The system-level
compilation and architectural implications of this approach have been
explored separately, where the continuous-transport model is shown
to be compatible with efficient circuit execution and resource management
\citep{Chang2025}. Together, these results indicate that the DTG
scheme provides a viable building block for scalable, modular trapped-ion
quantum computing architectures.

\begin{flushleft}
\textbf{Discussion }
\par\end{flushleft}

The DTG scheme builds upon the advantages of the QCCD architecture
while addressing some of its key drawbacks, making it a promising
alternative for scalable trapped-ion quantum computing. Similar to
the QCCD approach, the DTG scheme organizes ions into multiple arrays,
referred to as modules, and moves the communication ion close to a
specific module during entanglement operations. This design avoids
the need for scaling laser power with increasing inter-ion separation
during entanglement, maintaining efficiency even in large-scale systems.
However, unlike the QCCD architecture, the DTG scheme eliminates the
need for the communication qubit to stop for entanglement with stationary
ions. This continuous motion avoids motional heating caused by the
non-uniform acceleration and deceleration of ions, which is a significant
challenge in QCCD systems. Moreover, in the DTG framework, the communication
ions can be pre-cooled prior to the entanglement operation, eliminating
the need for employing different ion species for sympathetic cooling
during ion shuttling and thereby reducing system complexity and resource
overhead. The DTG scheme also offers flexibility for distributed quantum
operations, including superdense coding, quantum teleportation, and
the generation of multipartite entanglement. By avoiding the severe
motional heating associated with splitting and merging processes in
QCCD systems, the DTG scheme enables a more efficient and reliable
method for connecting distant ion modules. This capability provides
a pathway toward scalable modular trapped-ion architectures. 

In conclusion, we theoretically demonstrate and analyze a drive-through
entangling gate between a stationary ion and a mobile ion that remains
in continuous motion during the interaction. Our scheme simplifies
the transport protocol in existing QCCD architectures and is feasible
with current quantum technology. Precise control over ion positioning
has been demonstrated for transport gates \citep{Tinkey2022}, and
the Raman laser scheme for the adiabatic $\sigma_{z}\otimes\sigma_{z}$
gate has been well established for decades. By combining these techniques,
our approach provides a practical route for realizing entangling gates
in trapped-ion systems based on continuous ion transport.

Future work could further refine this scheme to reduce residual errors
and explore its integration into larger modular trapped-ion architectures.
\begin{flushleft}
\textbf{Method }
\par\end{flushleft}

\textbf{Derivation of the transformed Hamiltonian.} We deal with a
Hamiltonian $\hat{H}(t)=\hat{H}_{0}(t)+\hat{H}_{1}(t)$ with 
\begin{eqnarray*}
\hat{H}_{0}(t) & = & \frac{1}{2}\left[\hat{p}^{2}+\omega^{2}(t)\hat{q}^{2}\right],\\
\hat{H}_{1}(t) & = & f(t)\hat{q},
\end{eqnarray*}
which is a simplified version but the result can be generalized to
incorporate the state-dependence as in the main text. Here, $\hat{q}$
and $\hat{p}$ are the position and momentum operators, $\omega(t)$
is the time-dependent frequency, and $f(t)$ is the driving force.
Note that we have set mass $m=1$ for simplicity. Define the unitary
transformation 
\begin{eqnarray*}
\hat{U}_{D} & = & \exp\left\{ \frac{i}{\hbar}\left[\dot{u}(t)\hat{q}-u(t)\hat{p}\right]\right\} 
\end{eqnarray*}
 where $u(t)$ satisfies the classical equation of motion of the oscillator
\[
\ddot{u}(t)+\omega^{2}(t)u(t)=f(t).
\]
 In this sense, $\hat{U}_{D}$ is a displacement operator whose displacement
follows the classical trajectory of the system. The unitary transformation
$\hat{U}_{D}$ transforms $\hat{H}(t)$ into 
\[
\tilde{H}(t)=\hat{U}_{D}\hat{H}(t)\hat{U}_{D}^{\dagger}-i\hbar\hat{U}_{D}\frac{\partial\hat{U}_{D}^{\dagger}}{\partial t}.
\]
 The first term is given by 
\begin{align*}
 & \hat{U}_{D}\hat{H}(t)\hat{U}_{D}^{\dagger}\\
= & \frac{1}{2}\left\{ \left[\hat{p}-\dot{u}(t)\right]^{2}+\omega^{2}(t)\left[\hat{q}-u(t)\right]^{2}\right\} +f(t)\left[\hat{q}-u(t)\right]\\
= & \frac{1}{2}\left[\hat{p}^{2}+\omega^{2}(t)\hat{q}^{2}\right]-\dot{u}(t)\hat{p}+\frac{1}{2}\dot{u}^{2}(t)-\omega^{2}(t)u(t)\hat{q}\\
 & +\frac{1}{2}\omega^{2}(t)u^{2}(t)+f(t)\hat{q}-f(t)u(t).
\end{align*}
The time derivative of $\hat{U}_{D}^{\dagger}$ is 
\begin{align*}
 & \frac{\partial\hat{U}_{D}^{\dagger}}{\partial t}\\
= & \frac{i}{\hbar}\exp\left[\frac{i}{2\hbar}u(t)\dot{u}(t)\right]\exp\left[\frac{i}{\hbar}u(t)\hat{p}\right]\\
 & \times\left[\frac{1}{2}\dot{u}^{2}(t)+\frac{1}{2}u(t)\ddot{u}(t)+\dot{u}(t)\hat{p}-\ddot{u}(t)\hat{q}\right]\\
 & \times\exp\left[-\frac{i}{\hbar}\dot{u}(t)\hat{q}\right].
\end{align*}
Thus, the second term is 
\begin{align*}
 & -i\hbar\hat{U}_{D}\frac{\partial\hat{U}_{D}^{\dagger}}{\partial t}\\
= & \exp\left[\frac{i}{\hbar}\dot{u}(t)\hat{q}\right]\left[\frac{1}{2}\dot{u}^{2}(t)+\frac{1}{2}u(t)\ddot{u}(t)+\dot{u}(t)\hat{p}-\ddot{u}(t)\hat{q}\right]\\
 & \times\exp\left[-\frac{i}{\hbar}\dot{u}(t)\hat{q}\right]\\
= & -\frac{1}{2}\dot{u}^{2}(t)+\frac{1}{2}u(t)\left[-\omega^{2}(t)u(t)+f(t)\right]+\dot{u}(t)\hat{p}-\ddot{u}(t)\hat{q}.
\end{align*}
Combining the two terms, we have 
\begin{align*}
 & \tilde{H}(t)\\
= & \frac{1}{2}\left[\hat{p}^{2}+\omega^{2}(t)\hat{q}^{2}\right]+\left[-\ddot{u}(t)-\omega^{2}(t)u(t)+f(t)\right]\hat{q}\\
 & +\left[-\frac{1}{2}f(t)u(t)\right]\\
= & \frac{1}{2}\left[\hat{p}^{2}+\omega^{2}(t)\hat{q}^{2}\right]-\frac{1}{2}f(t)u(t)\\
= & \hat{H}_{0}(t)+\hat{\theta}(t),
\end{align*}
where $\hat{\theta}(t)=-\frac{1}{2}f(t)u(t)$ is the geometric phase
term. This derivation can be generalized by allowing $f(t)$ and $u(t)$
to incorporate state-dependent operators. Under this extension, the
geometric phase term $\hat{\theta}(t)$ yields the $\sigma_{z}\otimes\sigma_{z}$
interaction described in the main text.

\bibliographystyle{naturemag}
\bibliography{dtg}
 
\end{document}